# Moving liquids with light: Photoelectrowetting on semiconductors

Steve Arscott†

*Institut d'Electronique, de Microélectronique et de Nanotechnologie (IEMN), The University of Lille, Cité Scientifique, Avenue Poincaré, 59652 Villeneuve d'Ascq, France*

†steve.arscott@iemn.univ-lille1.fr

**Liquid transport in microchip-based systems[1-7] is important in many areas such as Laboratory-on-a-chip[8], Microfluidics[9] and Optofluidics[10]. Actuation of liquids in such systems is usually achieved using either mechanical displacement[11] or via energy conversion e.g. electrowetting[12-15] which modifies wetting. However, at the moment there is no clear way of actuating a liquid using light. Here, by linking semiconductor physics[16] and wetting phenomenon[17] a brand new effect "photoelectrowetting" is demonstrated for a droplet of conducting liquid resting on an insulator-semiconductor stack. Optical generation of carriers in the space-charge region of the underlying semiconductor alters the capacitance of the insulator-semiconductor stack; the result of this is a modification of the wetting contact angle of the droplet upon illumination. The effect is demonstrated using commercial silicon wafers, both n- and p-type having a doping range spanning four orders of magnitude ($6\times10^{14}$-$8\times10^{18}$ cm$^{-3}$), coated with a commercial fluoropolymer insulating film (Teflon®). Impedance measurements confirm that the observations are semiconductor space-charge related effects. The impact of the work could lead to new silicon-based technologies in the above mentioned areas[8-10].**

Over a hundred years after pioneering work on electro-capillarity phenomena[18], electrowetting has undergone a recent renaissance via the use of microtechnology[12-15]. Briefly, when a droplet of conducting liquid, e.g. an electrolyte, is placed on a conductor coated with a thin layer of insulating material (Fig. 1) the wetting contact angle $\theta$ of the droplet can be modified by applying a voltage $V$ according to the Young-Lippmann equation[17]:

$$\cos\theta = \cos\theta_0 + C\frac{V^2}{2\gamma}$$

where $\theta_0$ is the wetting contact angle at zero bias, $C$ is the capacitance (per unit surface) of the insulating layer and $\gamma$ is the surface tension of the liquid droplet. Note that *C does not vary* with either applied voltage or illumination for the liquid-insulator-conductor (LIC) system and the effect is *symmetrical with voltage polarity*, i.e. $\theta_{-V} = \theta_{+V}$. The effect has led to applications[13] such as displays[19,20], microelectromechanical systems[21], tunable optics[22] and digital-microfludics[23,24] but still holds fundamental scientific challenges[15].

In contrast, if we now replace the conductor in Fig. 1 by a *semiconductor* to form a liquid-insulator-semiconductor (LIS) stack shown in Fig. 2a, space-charge effects in the semiconductor can now modify the electrowetting behaviour as $C$ can change with both applied voltage[16] *and* illumination[25]. The reason for this is that $C$ is now composed of two capacitances connected in series: $C_i$, the insulator capacitance and $C_s$ the *voltage and illumination dependent capacitance* of the space-charge region in the semiconductor (Fig. 2b); an effect much exploited for image sensing[26,27].

By borrowing well-known language from semiconductor physics[16] we should be able to make some simple predictions about the electrowetting behaviour of the LIS system by making the analogy with the well-known metal-insulator-semiconductor (MIS) system. Indeed it is well known that depletion layers form at electrolyte-semiconductor interfaces[28,29]; this effect has been exploited for commercial purposes[30,31] and electrolyte-semiconductor[32] and electrolyte-insulator-semiconductor[33] interfaces have been used for transistor-based sensing[34,35] and carbon nanotube transistors[36].

Straight away one can predict major differences compared to the LIC system: the electrowetting behaviour (i) will be *asymmetrical* with voltage polarity ($\theta_{-V} \neq \theta_{+V}$) (ii) will depend on the doping *density* and *type* of the underlying semiconductor and (iii) will depend on *illumination* (the value of $C_s$, can be modified by incident above band gap radiation)[16,25], (iv) under accumulation or high doping density should behave as the regular LIC system and (v) will be governed by other semiconductor effects, e.g. carrier inversion and breakdown[16] at higher bias.

Fig. 2c shows the predicted effect of illumination using above band gap light on a biased LIS system in *depletion*. Light penetrating into the semiconductor will generate electron-hole pairs in the depletion zone[16]; the effect of this is to *increase* the capacitance of the stack[25] and thus *reduce* the contact angle of the droplet *at constant voltage*, i.e. the droplet spreads out upon illumination. Provided that there is no electrochemistry during illumination, e.g. due to breakdown of the insulator or the semiconductor, then the effect should be reversible.

The simple arguments above are based on the following assumptions: (i) the Debye length $\lambda_D$ in the electrolyte (see Supplementary Information) is much less than the thickness of the insulator and the depletion width in the semiconductor meaning electrical double layer (EDL) effects[32,33] need not be considered, (ii) the capacitance is uniform under the droplet[37] and (iii) interface effects are ignored for the moment (surface states, trapped charge, mobile ionic charge in the insulator, high field effects…)[16].

In order to test the above ideas commercial 3" silicon wafers (Siltronix, France) were coated with a fluoropolymer layer of polytetrafluoroethylene (PTFE). PTFE is commercially available as Teflon® AF (Dupont, USA). Fluoropolymer films are currently being investigated for transistor gate dielectric[38,39]. The Teflon® thicknesses were 265 nm (±15 nm) and 20 nm (±3 nm) using spin coating (see Supplementary Information). Four types of silicon wafers were employed: p-type ($N_A \sim 1.8 \times 10^{15}$ cm$^{-3}$ and $N_A \sim 8 \times 10^{18}$ cm$^{-3}$) and n-type ($N_D \sim 6 \times 10^{14}$ cm$^{-3}$ and $N_D \sim 3.5 \times 10^{17}$ cm$^{-3}$) (see Supplementary Information); these will be referred to as p+, p, n+ and n henceforth. Aluminium Ohmic contacts were formed on the rear surface of the wafers (see Supplementary Information).

The contact angle data was gathered using a commercial Contact Angle Meter (GBX Scientific Instruments, France) (see Supplementary Information) and the experiments were performed in a class ISO 5/7 cleanroom ($T = 20°C \pm 0.5°C$; $RH = 45\% \pm 2\%$).

Five liquids were used for the experiments: (i) a dilute hydrochloric acid solution (concentration = 0.01M) (ii) two saline solutions (0.01M and 0.001M) (iii) an acetic acid solution (1M) and Coca-Cola zero® (The Coca-cola Company, USA). The electrical conductivities of the solutions were measured to be 3.64 mS cm$^{-1}$, 1.18 mS cm$^{-1}$, 0.12 mS cm$^{-1}$, 1.32 mS cm$^{-1}$ and 1.07 mS cm$^{-1}$ respectively (see Supplementary Information).

A small volume (~µL) of liquid was placed on the Teflon® surface. One contact was provided using the Ohmic contact on the rear surface of the wafers. The voltage (±40V) was applied to the liquid via a stainless steel wire inserted into the droplet and the contact angle was recorded (photographed and filmed) (see Supplementary Information). The applied voltage was stepped to the measurement voltage $V$ (see Supplementary Information).

Fig. 3 shows photographs of the droplet profile as a function of applied voltage for the different silicon wafers under ambient room lighting (<10 W m$^{-2}$). For the two higher doped silicon wafers (p+ and n+) the contact angle measurements did not reveal a marked asymmetric contact angle variation with voltage polarity (Figs. 3a and 3b) behaving very much as the LIC system i.e. $\theta_{-V} \sim \theta_{+V}$. In contrast, in the case of the two lower doped silicon wafers (p and n) a distinct electrowetting asymmetry, depending on the voltage polarity, is observed (Figs. 3c and 3d). If we consider the p-type sample, a negative voltage leads to a contact angle variation similar to the p+ sample (Fig 3c); however, for a positive voltage the contact angle variation is less, i.e. the droplet tends not to spread out for positive applied voltages. The opposite is observed for the n-type silicon in terms of voltage polarity (Fig. 3d), i.e. the droplet tends not to spread out for negative applied voltages.

Interestingly asymmetrical electrowetting[40] was reported for a silicon wafer (doping type and density not given in the paper) covered with a thin layer of Teflon® although this was attributed at the time to preferential adsorption of hydroxide ion, OH⁻ at Teflon® AF-water interfaces. The results here involving the highly doped silicon wafers (p+ and n+) point towards a semiconductor related effect rather than a Teflon®-liquid interaction.

Fig. 4 shows the effect of illumination on the droplet profile for the different silicon wafers and liquids. Under depletion conditions, a negative voltage for n-type and a positive voltage for p-type, illumination with an above band gap white light source (Schott, USA) (see Supplementary Information) leads to the expected flattening out of the droplet at constant voltage, i.e. the predicted photoelectrowetting effect. The photoelectrowetting effect is reversible (see Supplementary Films), switching off the light source causes the droplet to "pop-up" and the dark contact angle is restored; the process can be repeated many times. As the voltage is increased on the droplet (positive for p-type, negative for n-type) the droplet flattens out more upon illumination, i.e. the illuminated contact angle ($\theta_{Light}$) decreases with increasing voltage.

Fig. 5 shows plots of the droplet contact angle as a function of applied voltage and illumination for the different silicon wafers. The two highly doped silicon wafers (p+ and n+) follow a classic relationship as expected from the Young-Lippmann equation (dashed black line) with $C$ near constant. In contrast, the electrowetting for the lower doped p- and n-type silicon wafers does not at all follow a classic relationship for a positive and negative applied voltage respectively. In the case of the n-type wafer coated with 265 nm of Teflon® the contact angle at -30V reduces by ~7° whereas at +30V it has reduced by ~20°. For the p-type wafer coated with 265 nm of Teflon®, the contact angle at -30V reduces by ~26° whereas at +30V it has reduced by ~14°. The continuous coloured curves are based on results obtained during the electrical characterization (see later).

Fig. 5 also shows the variation of droplet contact angle with illumination for the two lower doped silicon wafers (n and p). The largest optically induced contact angle change $\Delta\theta_L = \theta_{Dark} - \theta_{Light}$ equal to ~16° is observed for the n-type silicon at -40V (Fig. 5b) although by reducing the thickness of the Teflon® from 265 nm to 20 nm on the n-type wafer a $\Delta\theta_L$ ~15° can be achieved at voltages around 10V (Fig. 5c). The green arrows indicate reversibility.

The films (see Supplementary Information) enable a comparison of the time it takes the contact angle to adjust from $\theta_{Dark}$ to $\theta_{Light}$ and from $\theta_0$ to $\theta_{\pm V}$ in the absence of illumination.

These times were found to be comparable (~70 ms) supporting the view that the effect of light is to switch from a depleted state to an illuminated state[16,25] in the semiconductor via the generation of electron-hole pairs in the space-charge region. Additional experiments were preformed using bare silicon wafers not coated with Teflon® where a Schottky barrier is known to form[28,29]; a non-reversible photoelectrowetting effect was observed at low voltages (see Supplementary Information).

In terms of optically induced liquid transport (thermal effects[11] aside), the literature is scarce[41-44]. Laser-induced microflow due to vortices is possible[41,42] and an optically activated electrowetting has been reported[43] but was due to a photoconductor used as a *lumped circuit element* rather than an intrinsic space-charge layer related effect which can be associated with the capacitance in the Young-Lippmann equation as is reported here. A "photo wetting" effect has also been reported using liquid crystals[44] but the physical mechanism responsible for the observations is not yet clear.

All this is well and good but in order to provide evidence that the photoelectrowetting and asymmetrical electrowetting effects observed here are related to *space-charge effects in the semiconductor* we need to look into what is happening *underneath* the surface; this can be done using impedance measurements.

Gold contacts (surface area ~ 0.1-1 mm$^2$) were evaporated onto the Teflon®-silicon samples through a mechanical mask to form MIS structures. Impedance measurements were gathered using a Precision Impedance Analyzer (Agilent, USA) (see Supplementary Information).

Fig. 6 shows the capacitance-voltage curves for the gold-Teflon®-silicon stacks. For the highest doped (p+) silicon wafer (black squares) the capacitance is constant [$6.43 \times 10^{-5}$ F m$^{-2}$ ($\pm 0.004 \times 10^{-5}$ F m$^{-2}$)] over the voltage range ($\pm 20$V) (Fig. 6a). The low frequency dielectric constant of the Teflon® can be extract from this to be 1.92 ($\pm 0.04$). For the next highest doped wafer (n+) (black circles), a slight carrier accumulation/depletion effect is observed with the capacitance varying from $6.26 \times 10^{-5}$ F m$^{-2}$ to $6.4 \times 10^{-5}$ F m$^{-2}$ as the voltage is increased from -20V to +20V (Fig 6a). Illumination using the white light source was observed to have little effect on the capacitance of these two samples just as illumination had no noticeable effect on the contact angle during the photoelectrowetting experiments for these samples (p+ and n+).

In terms of the lower doped samples (p and n), Fig 6b and Fig 6c show the variation of capacitance with voltage under background room lighting <10 W m$^{-2}$ (black squares). In this case the capacitance is observed to vary strongly with voltage as the depletion width in the semiconductor is modified. The stacks behave similar to MIS system, for the n-type wafer coated with 265 nm of Teflon® the capacitance is reduced from $6.14 \times 10^{-5}$ F m$^{-2}$ to $1.2 \times 10^{-5}$ F m$^{-2}$ as the applied voltage is reduced from 0V to -40V. In the case of the p-type wafer coated with 265 nm of Teflon® the capacitance reduces to $3.1 \times 10^{-5}$ F m$^{-2}$ from $6 \times 10^{-5}$ F m$^{-2}$ as the voltage is increased from -10V to +40V.

Figs. 6b and 6c show the effect of illumination on the capacitance-voltage curves of the lower doped silicon wafers using above band gap light using the same white light source (Schott, USA) as used for the electrowetting experiments. As expected, the effect of illumination is to increase of the capacitance of the stacks[25]. It is this increase in capacitance that causes the droplet to spread out in the electrowetting experiments upon illumination. Indeed, the measured capacitance-voltage curves (dark and light) can now be inserted into the Young-Lippmann equation to compare the electrical characterization with the electrowetting characterization. These are the continuous coloured curves visible on Fig. 5. The capacitance-voltage variations for the highly doped wafers (blue curves in Fig. 5) follow the Young-Lippmann equation with $C$ near constant. The capacitance-voltage variations for the lower doped wafers (red curves in Fig. 5) do not follow the Young-Lippmann equation as $C$ varies with $V$. The capacitance variation observed during the impedance measurements under illumination fits very well with the contact angle variation upon illumination during the photoelectrowetting experiments (pink curves in Fig. 5). It is very clear that the electrical characterization gives very strong evidence that the asymmetrical electrowetting and photoelectrowetting observations here are semiconductor space-charge related effects[16].

It was observed that the light intensity (or more correctly the *irradiance* W m$^{-2}$) required to vary the capacitance (during the impedance measurements) was greater than was necessary to produce the equivalent photoelectrowetting response by a factor of ~2-3. In order to understand why this is so the optical transmission into the silicon for the air-water-Teflon®-silicon and air-gold-Teflon®-silicon stacks was calculated (see Supplementary Information). The generation rate $g$ cm$^{-3}$ s$^{-1}$ of photogenerated carriers in the silicon could thus be estimated and compared to an established photocapacitance model for MIS capacitors[25]. This calculation was in good agreement with observations and showed that (i) $g_{\text{photoelectrowetting}}/g_{\text{impedance}}$ ~3 for the same irradiance and (ii) $g_{\text{photoelectrowetting}} > 10^{20}$ cm$^{-3}$ s$^{-1}$ to produce the observed angle changes in the photoelectrowetting experiments. The photocapacitance model[25] yields a value of this order (see Supplementary Information). Going further, the MIS photocapacitance model[25] can be injected into the Young-Lippmann equation to calculate the variation of contact angle as a function of electron-hole pair generation rate $g$ in the silicon (see Supplementary Information). The results of this are given in the following table and compare well with the experimental photoelectrowetting results given in Fig. 4.

| Depletion Voltage | p-type Measured | Model[25] | n-type Measured | Model[25] |
|---|---|---|---|---|
| 30 | 3.8° | 6.5° | 9.4° | 9.2° |
| 35 | 10° | 8.7° | 9.7° | 12.3° |
| 40 | 12.8° | 11.4° | 16.5° | 16° |

**Table**: Measured and modelled contact angle variation under illumination. The measured values are for silicon coated with 265 nm of Teflon® using a 0.01M HCl solution. The calculated values use the photocapacitance model[25] injected into the Young-Lippmann equation.

We are now in a position to understand the electrowetting and photoelectrowetting on semiconductors in terms of energy bands. Fig. 7 illustrates the role of the carriers in the semiconductor in electrowetting and photoelectrowetting on a semiconductor coated with an insulator. If we take the example of an underlying p-type semiconductor; for a negative applied voltage then *accumulation* of carriers (holes) can occur near the insulator/semiconductor interface (Fig. 7a)[16]; in this case the electrowetting behaviour will behave as the LIC system ($C = C_i$) and the droplet will spread out. If now we apply a positive voltage to the droplet, *depletion* of carriers (holes) will occur near the insulator/semiconductor interface and $C_s$ will diminish with increasing voltage[16] resulting in less spreading out of the droplet (Fig. 7b) as would occur at the equivalent negative voltage. If the positive bias is further increased then depending on doping level of the semiconductor carrier inversion (electrons) could occur[16] at the insulator/semiconductor interface leading to a re-emergence electrowetting as in the LIC system (Fig. 7c). Under depletion, illumination using above band gap light generates electron-hole pairs in the space-charge region leading to an increase of the overall capacitance (Fig. 7d) and spreading out of the droplet, i.e. photoelectrowetting.

As a last point, it is important to understand that the effects described here are *not* the result of simply adding a lumped element into the circuit[43] which acts as an optically controlled switch. On the contrary, the depletion layer in the underlying semiconductor is intrinsically responsible for the electrowetting and photoelectrowetting behaviour via the Young-Lippmann equation. Thus, the electrowetting and photoelectrowetting here is governed by the *bulk* electronic properties of the underlying semiconductor. The observations indicate a previously uncharted overlap between *capillary phenomena* and *semiconductor physics* which may prove to be valuable for the introduction of semiconductor effects into the above mentioned technologies[8-11] where wetting and capillary effects are fundamental.

**Supplementary Information** The supplementary Information contains experimental details necessary in order to repeat the study. Films are also provided which clearly demonstrate the optically activated electrowetting reported here.

**Acknowledgements** The author would like to thank Gérard Cambien for technical help with the liquid resistivity measurements and Etienne Okada for technical help with impedance measurements.

**Figure 1. Electrowetting behaviour for the liquid-insulator-conductor system.** The variation of the contact angle of the droplet is symmetrical with the applied voltage polarity, $\theta_{-V} = \theta_{+V}$, and obeys the Young-Lippmann equation. The capacitance of the insulating layer is $C$ Fm$^{-2}$; the conductor is considered to be perfect; the liquid is considered to be conducting.

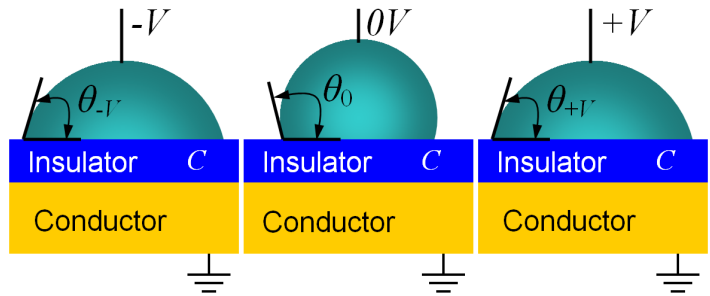

**Figure 2. Electrowetting and photoelectrowetting behaviour for the liquid-insulator-semiconductor system**. **a**, The variation of the contact angle is not symmetrical with voltage polarity, $\theta_{-V} \neq \theta_{+V}$, due to the formation of a depletion region in the semiconductor under the insulator. The capacitance of the insulating layer is $C_i$ Fm$^{-2}$; the voltage dependent capacitance of the semiconductor is $C_s$ Fm$^{-2}$. **b**, Effect of illumination. Contact angle at positive bias ($\theta_{Dark}$) for a p-type semiconductor and optically modified contact angle ($\theta_{Light}$) at the same positive bias via the formation of photogenerated carriers (electron-hole pairs) in the semiconductor. The green arrow represents illumination using above band gap light. The arrows indicate that the effect is reversible. The photoelectrowetting angle modification $\Delta\theta_L = \theta_{dark} - \theta_{Light}$. **c**, Voltage and illumination dependent capacitances involved in electrowetting and photoelectrowetting.

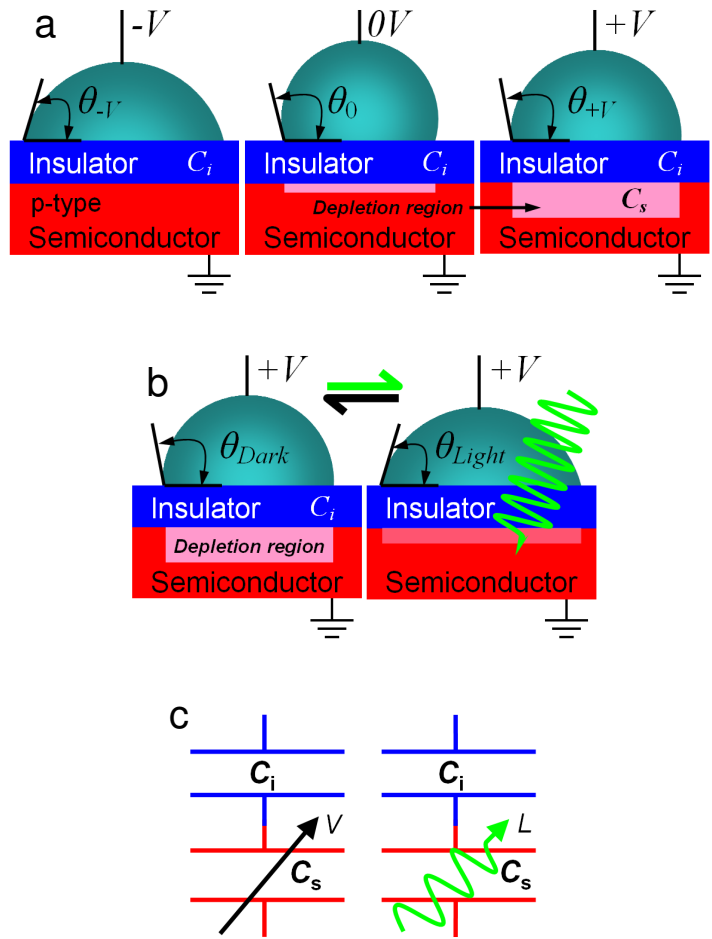

**Figure 3. Droplet profiles for electrowetting on Teflon® coated silicon wafers**. **a**, p+ type. **b**, n+ type. **c**, p type and **d**, n type  All silicon wafers were coated with a 265 nm thick layer of Teflon® AF. Liquid = HCl solution (0.01M).

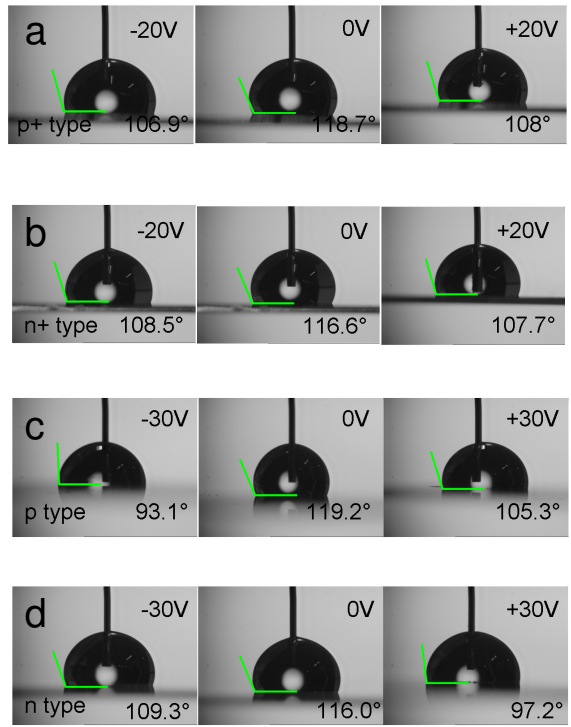

**Figure 4. The effect of illumination on droplets resting on Teflon® coated silicon wafers**. **a**, p-type silicon coated with 265 nm of Teflon® using and HCl solution (0.01M). **b**, p-type silicon coated with 265 nm of Teflon® using and Acetic acid solution (1M). **c**, n-type silicon coated with 265 nm of Teflon® using and HCl solution (0.01M). **d**, n-type silicon coated with 265 nm of Teflon® using Coca-Cola Zero®. **e**, n-type silicon coated with 20 nm of Teflon® using and NaCl solution (0.001M). $\Delta\theta_L = \theta_{Dark} - \theta_{Light}$. Illumination was achieved using a white light source (Schott, USA).

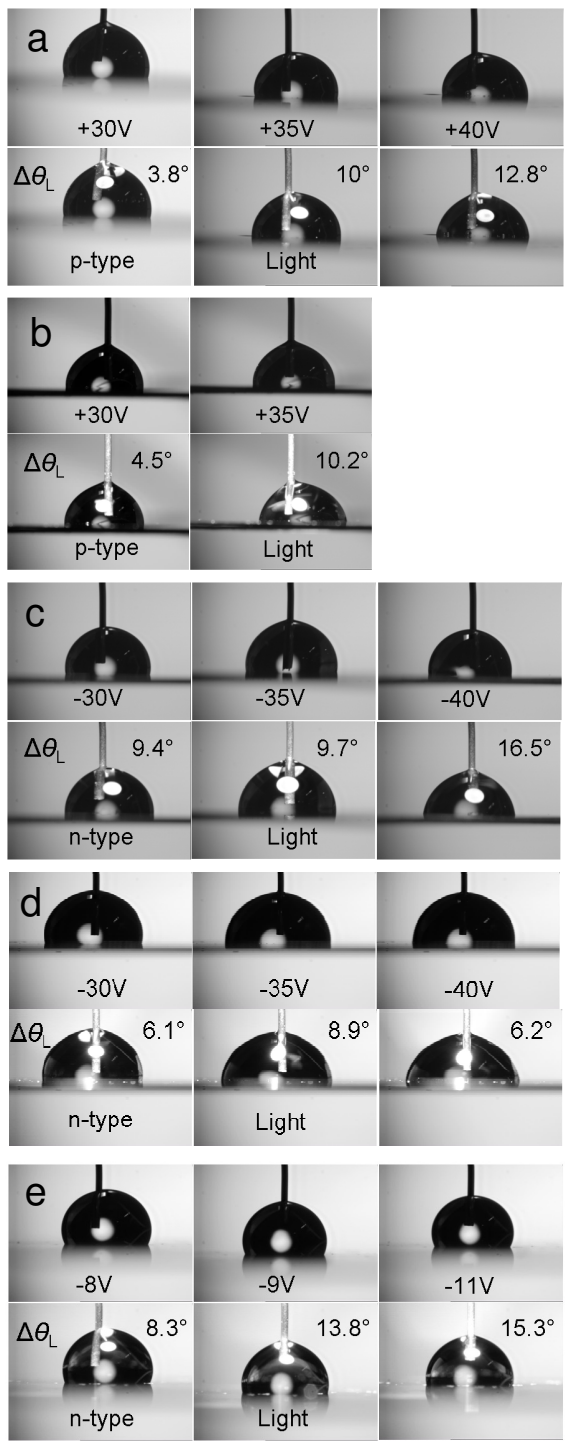

**Figure 5. Variation of droplet contact angle as a function of applied voltage and illumination for different silicon wafers coated with Teflon®**. **a**, p+ (blue circles) and p-type silicon (red circles) coated with 265 nm of Teflon®. **b**, n+ (blue circles) and n-type silicon (red circles) coated with 265 nm of Teflon®. The contact angles under illumination are indicated by pink circles. **c**, n+ (blue triangles) coated with 20 nm of Teflon® using an HCl (0.01M) solution, n-type silicon (open red squares) coated with 20 nm of Teflon® using an HCl (0.01M) solution, p-type silicon (open blue circles) coated with a 250 nm $SiO_2$ and 20 nm of Teflon® using an HCl (0.01M) solution and n-type silicon (open green squares) coated with 20 nm of Teflon® using an NaCl (0.001M) solution. The effect of illumination (open pink squares); the green arrows indicate reversibility. The dashed black lines correspond to the Young-Lippmann equation. The solid lines (red, blue and pink) correspond to the capacitance-voltage results (see later) injected into the Young-Lippmann equation. Room lighting measured to be <10 W $m^{-2}$. Illumination: p-type – 570 W $m^{-2}$, n-type - 1250 W $m^{-2}$.

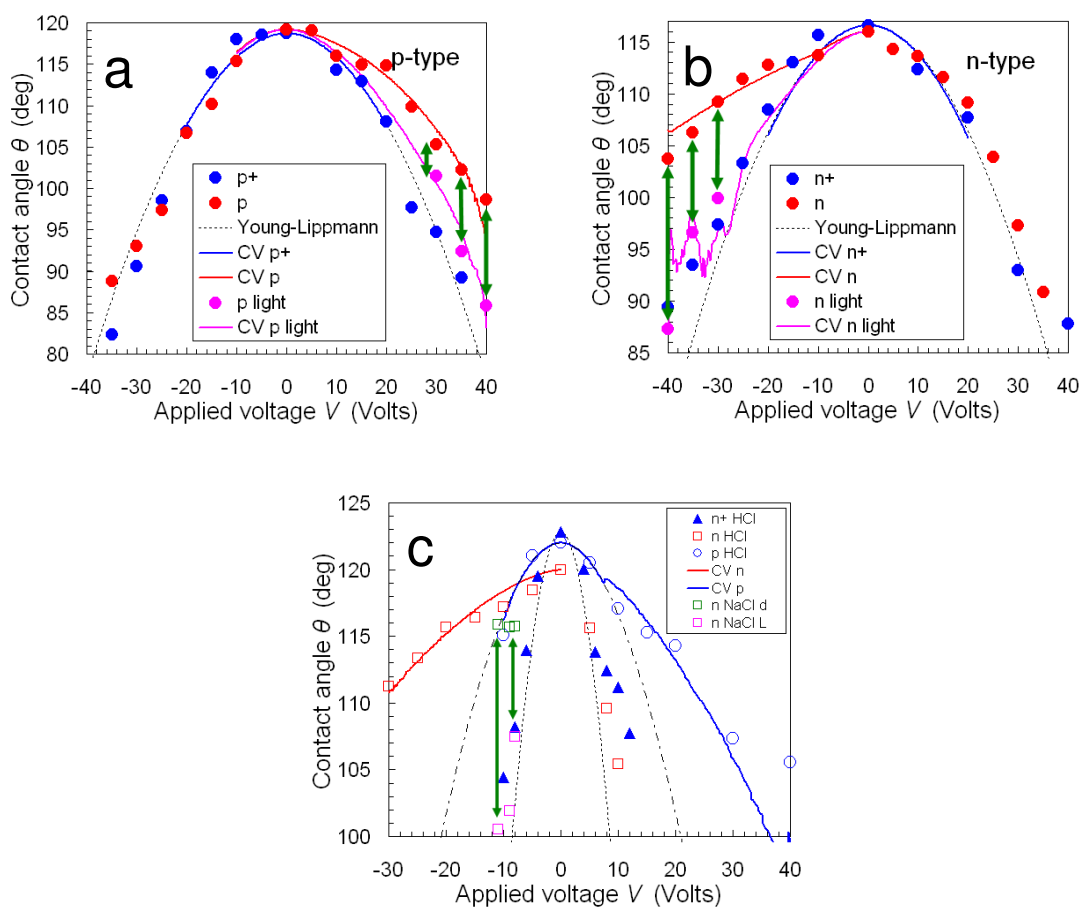

**Figure 6. Capacitance-voltage profiles of gold-Teflon®-silicon stacks**. **a,** p+ and n+ silicon wafers coated with 265 nm of Teflon®. **b**, p-type silicon coated with 265 nm of Teflon®. **c**, n-type silicon coated with 265 nm of Teflon®. The coloured squares: red to green correspond to increasing light intensity. p-type: red = 350 W m$^{-2}$, orange = 500 W m$^{-2}$, yellow = 700 W m$^{-2}$ and green = 1100 W m$^{-2}$. n-type: red = 200 W m$^{-2}$, orange = 1200 W m$^{-2}$, yellow = 2000 W m$^{-2}$ and green = 2600 W m$^{-2}$. The black traces correspond to the measurement under room lighting <10 W m$^{-2}$. Measurement frequency for **a** = 10 kHz, **b** = 100 kHz, **c** = 10 kHz.

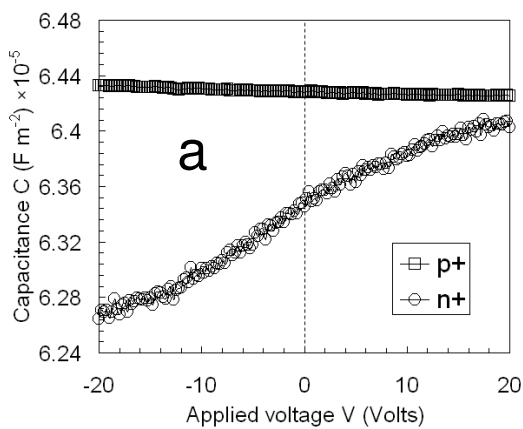

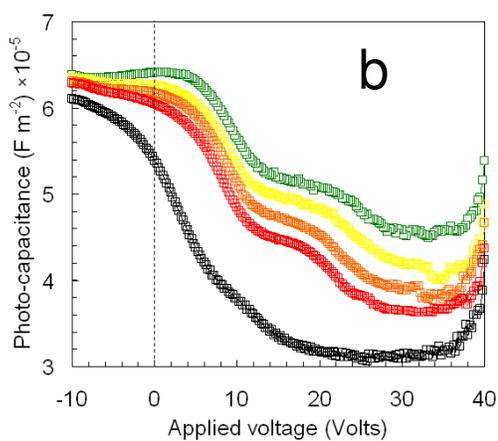

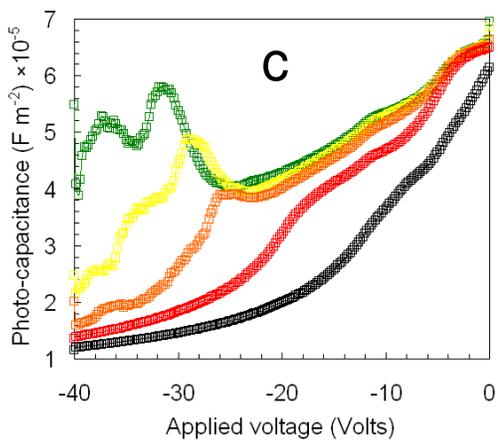

**Figure 7. The role of energy bands and carriers in electrowetting and photoelectrowetting for p-type silicon coated with an insulator**. **a**, accumulation, **b**, depletion, **c**, inversion and **d**, illumination with above band-gap light under depletion. The insulator is light blue. The green arrow indicates illumination with above band-gap light; the blue arrow corresponds to electron-hole pair photogeneration. $E_v$, $E_F$, $E_i$ and $E_c$ correspond to the valence band energy, the Fermi energy, the intrinsic Fermi energy and the conduction band energy in the semiconductor.

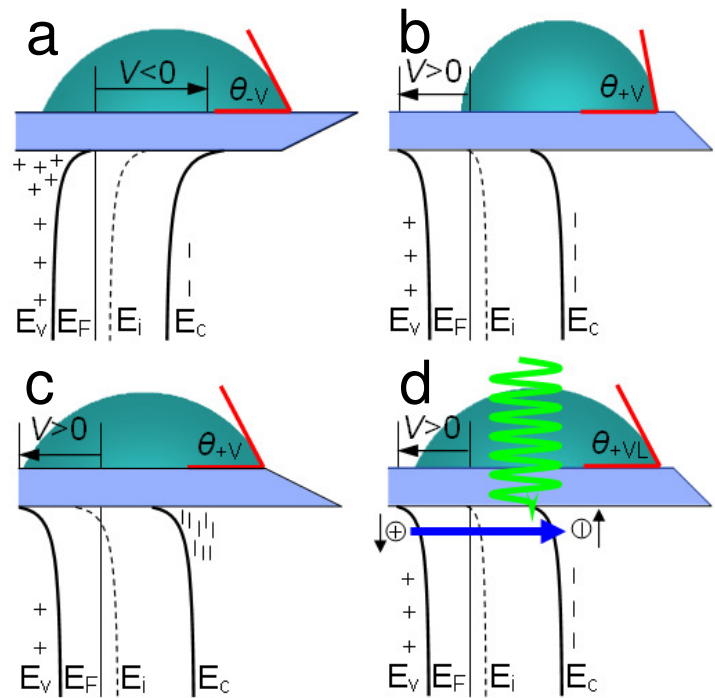

# Supplementary Information for "Moving liquids with light: Photoelectrowetting on semiconductors".

**Steve Arscott**†
*Institut d'Electronique, de Microélectronique et de Nanotechnologie (IEMN), The University of Lille, Cité Scientifique, Avenue Poincaré, 59652 Villeneuve d'Ascq, France*
†steve.arscott@iemn.univ-lille1.fr

## A. Silicon wafers.

### 1. Wafer supply.

The silicon wafers were purchased from Siltronix, France (http://www.siltronix.com/).

### 2. Wafer data.

Information on silicon wafers is given in Supplementary Tables I and II:

| wafer | Species | Resistivity (Ω cm) | Average Doping (cm$^{-3}$) |
|---|---|---|---|
| n-type | P | 5-10 | $6 \times 10^{14}$ |
| p-type | B | 5-10 | $1.8 \times 10^{15}$ |
| n+type | P | 0.03-0.05 | $3.5 \times 10^{17}$ |
| p+type | B | 0.009-0.01 | $8 \times 10^{18}$ |

**Supplementary Table I**: silicon wafers used for the study (electrical data)

| wafer | diameter | Thickness (μm) | Polished rear side | Orientation |
|---|---|---|---|---|
| n-type | 3" | 380±25 | no | (100) |
| p-type | 3" | 380±25 | no | (100) |
| n+type | 3" | 380±25 | no | (111) |
| p+type | 3" | 380±25 | no | (100) |

**Supplementary Table II**: silicon wafers used for the study (properties)

## B. Fabrication of samples.

### 1. Wafer preparation.

Prior to spin-coating deposition of the Teflon® AF 1600 (Dupont, USA) layer the silicon wafers were cleaned in the following way. $H_2SO_4/H_2O_2$ vol/vol = 3/1 for 1 minute followed by HF 50% for 30 seconds and drying by dry nitrogen. The wafers were then dehydrated at 200°C for 1 hour as the presence of water can cause problems with polymer dielectrics[S1].

[S1]Noh, Y. H., Park, S. Y., Seo, S-M., & Lee, H. H. Root cause of hysteresis in organic thin film transistor with polymer dielectric *Organic Electronics* **7**, 271–275 (2006).

2. **Teflon® AF deposition.**

The deposition parameters for the Teflon® AF 1600 are given in Supplementary Table III. The thickness of the Teflon® films was measured using surface profiling techniques.

| | | |
|---|---|---|
| **thickness** | **20 nm (±3 nm)*** | **265 nm (±15 nm)*** |
| Dilution of stock solution | 0.3% using Fluorinert FC-75 (3M, USA) | 2% using Fluorinert FC-75 (3M, USA) |
| Spin-coating | 1000rpm/1000rpms$^{-1}$/30 s | 1000rpm/1000rpms$^{-1}$/30 s |
| Bake1 | 50°C/5 min | 50°C/5 min |
| Bake2 | 110°C/5 min | 110°C/5 min |
| Bake3 | 170°C/5 min | 170°C/5 min |
| Bake4 | 330°C/15 min | 330°C/15 min |

**Supplementary Table III**: Teflon® AF 1600 deposition parameters. *Measured by surface profiling. The thicknesses agree well with Moon et al[40].

3. **Fabrication of the Ohmic contacts to the rear surfaces.**

Ion implantation was used to heavily dope the rear surface of the wafers to $N \sim 10^{20}$ cm$^{-3}$ (uniformly over a depth of ~100 nm) using boron (for the p-type wafers) and phosphorous (for the n-type wafers). The doping species were activated using high temperature rapid annealing and following aluminium deposition were annealed in an $N_2/H_2$ to activate the Ohmic contacts. The specific contact resistivity of the contacts was determined to be <$10^6$ $\Omega$ cm$^2$ using evaporated resistance ladders.

## C. Electrowetting experiments.

1. **Electrical set-up + light source.**

Supplementary Fig. 1 shows the set-up for the experimental experiments.

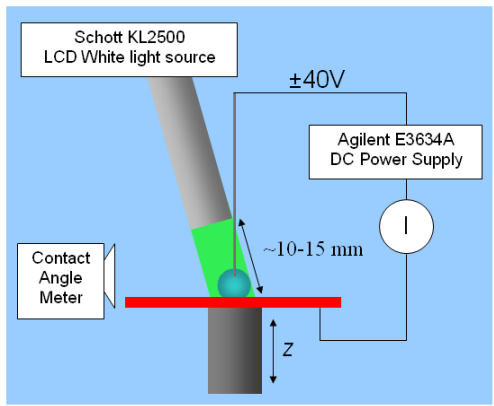

**Supplementary Figure 1**: Experimental set-up for the electrowetting experiments.

The experimental procedure for the electrowetting experiments was as follows: (i) a droplet of liquid (~1.5-1.8 µL) is placed on the Teflon® coated silicon wafer (red) using a pipette. (ii) a stainless steel wire (φ~150 µm) is dipped into the droplet (the contact angle meter allows observation of this). (iii) The applied DC voltage magnitude and polarity is varied *in distinct steps*. (iv) The droplet profile is recorded (photographed and filmed) using a DigiDrop Contact Angle Meter (GBX, France). (v) The effect of illumination on the droplet profile is recorded (photographed and filmed). The current is monitored to observed breakdown/electrolysis effects. Several measurements were performed for each wafer type; in general the contact angle results are accurate to ±1°.

The white light source (Schott, USA) uses a halogen reflector lamp Type ELC. In order to calibrate the light source a broadband Power Meter 210 (Coherent, USA) was employed. Thirty settings are available on the white light source A1-E6, the out power varies from 7.5 mW (A1) to 825 mW (E6).

In terms of the irradiance $I$ (W m$^{-2}$) of the light source (Schott, USA), the broadband power $P$ (W) was accurately determined the Power Meter (Coherent, USA). The output of the white light source is a 4mm diameter optical fibre bunch; this was placed at a distance $d$ of 10-15 mm from the droplets (electrowetting experiments) and gold contacts (impedance measurements). If we consider a Gaussian beam then the average irradiance $I$ incident on the samples can be approximated by $P/\pi d^2$; these are the values given in the paper and used in the calculation of the electro-hole pair generation rate.

Note that the photoelectrowetting films indicate that the time for the droplet to readjust to dark conditions is longer than the adjustment to illuminated conditions (~70 mS). This is due to the long relaxation time of the light source (>70 mS) when turned off.

The absorbance of the Teflon® layer is not considered here as Teflon® AF films are virtually transparent to white light[S2] as is water[S3].

[S2]Yang, M. K., French, R. H. & Tokarsky, E. W. Optical properties of Teflon® AF amorphous fluoropolymers. *J. Micro/Nanolith. MEMS MOEMS* **7**, 033010-9 (2008).
[S3]Sullivan S.A. Experimental Study of the Absorption in Distilled Water, Artificial Sea Water, and Heavy Water in the Visible Region of the Spectrum. *J. Opt. Soc. Am.* **53**, 962-967 (1963).

However, a calculation of the light entering the silicon layer for the (i) air-water-Teflon®-silicon stack (electrowetting experiments) and (ii) the air-gold-Teflon®-silicon (photocapacitance measurements) is possible[S4-S6]. For the electrowetting experiments ~60% of the light is entering the silicon layer whereas for the photocapacitance measurements around 20% of the light is entering the silicon samples.

[S4]Heavens, O. S. Optical properties of thin films. Rep. Prog. Phys. **23** 1-65 (1960).
[S5]Tomlin, S. G. Optical reflection and transmission formulae for thin films.brit. J. Appl. Phys. (J. Phys. D) **1**, 1667-1671 (1968).
[S6]Tomlin, S. G. More formulae relating to optical reflection and transmission by thin films. J. Phys. D: Appl. Phys. **5**, 847-851 (1972).

## 2. Solutions used for the measurements

**Solution conductivity**

The conductivity of the liquids was measured using a Radiometer Copenhagen CDM-83 Conductivity Meter. A calibration measurement was performed using a KCl (0.1M) solution prior to tests. The resistivity of the deionised water was >10 MΩ cm. Note that the Coca-cola Zero® was left in an open container for 7 days prior to measurements.

**Debye lengths in the solutions**

The Debye length $\lambda_D$ of the solutions was calculated using the following equation:

$$\lambda_D = \sqrt{\frac{\varepsilon_0 \varepsilon_r kT}{2q^2 c N_A}}$$

and the following numbers: $\varepsilon_r = 80$, $k = 1.38 \times 10^{-23}$ JK$^{-1}$, $T = 300$, $q = 1.6 \times 10^{-19}$ C, $c$ = solution molarity mol m$^{-3}$ and $N_A = 6.02 \times 10^{23}$ mol$^{-1}$. For a 0.01M (mol L$^{-1}$) HCl solution, $\lambda_D \sim 3$ nm and for a 0.001M saline solution, $\lambda_D \sim 9$ nm.

**Debye lengths in the silicon**

For a semiconductor the Debye length is given by the following equation[16]:

$$\lambda_D = \sqrt{\frac{\varepsilon_0 \varepsilon_r kT}{q^2 N}}$$

This enables the computation of the following table:

| Doping density $N$ (cm$^{-3}$) | Debye length $\lambda_D$ (nm) |
|---|---|
| 6×10$^{14}$ | 168.5 |
| 1.8×10$^{15}$ | 97.3 |
| 3.5×10$^{17}$ | 7 |
| 8×10$^{18}$ | 1.5 |

**Supplementary Table IV**: Debye lengths in the silicon wafers.

## 3. Drop shape analysis and contact angle extraction.

Once the droplet profiles have been recorded (photographs and films) the data can be analysed using appropriate software (Windrop++) available with the DigiDrop Contact Angle Meter (GBX, France). A freeware software was also used to extract the contact angles[S7].

[S7]Stalder A.F., Kulik G., Sage D., Barbieri L., & Hoffmann P. A Snake-Based Approach to Accurate Determination of Both Contact Points and Contact Angles. Colloids And Surfaces A: Physicochemical And Engineering Aspects **286**, 92-103 (2006).

## D. Electrical characterization.

1. **Fabrication of samples**

Titanium/Gold (Ti/Au, 10 nm/100 nm) were formed on the surface of the Teflon® using electron beam evaporation techniques using a metal mask which contained prefabricated holes. The thin titanium layer acts as an adherence layer for the gold to the Teflon®. The contact surfaces ranged from 0.1 to 1 mm$^2$. For the photocapacitance measurements a 20 nm thick semitransparent gold layer was deposited using a metal mask.

2. **Capacitance-Voltage (C-V) Measurements.**

The C-V measurements were carried out using a Precision Impedance Analyzer 4294A (Agilent, USA) using a bias voltage of ±40V. A full calibration (open circuit – load (200Ω) – short circuit) was performed using a P/N101-190 S/N33994 Impedance Standard Substrate (Cascade Microtech, USA) over the frequency range (500 Hz – 1MHz) prior to the measurements.

The experimental set-up for these experiments is shown in Supplementary Figure 2.

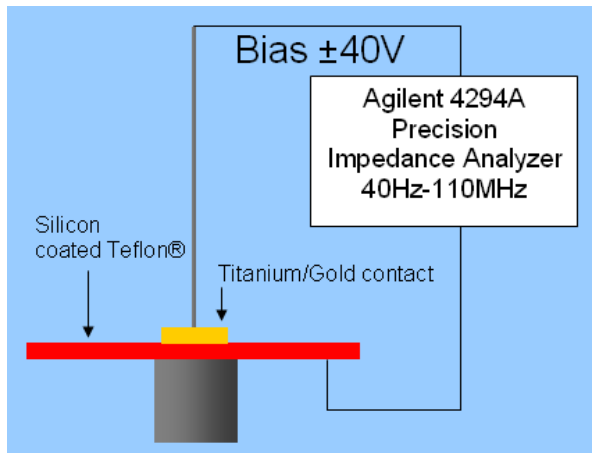

**Supplementary Figure 2**: Experimental set-up for the impedance experiments.

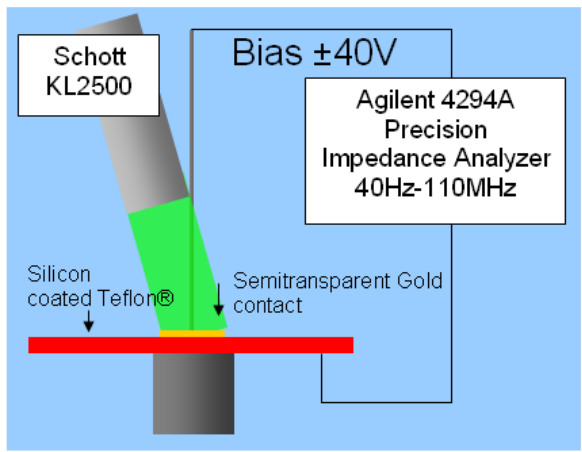

**Supplementary Figure 3**: Experimental set-up for the photo-capacitance experiments.

## 3. Model for the depletion capacitance

In depletion mode[16], the capacitance per unit area $C$ of an ideal conductor-insulator-semiconductor system, e.g. a metal-oxide-semiconductor (MOS), can be described by the following equation:

$$C = \frac{C_i}{\sqrt{1 \pm (2C_i^2 V / \varepsilon_s \varepsilon_0 q N)}}$$

where $C_i$ is capacitance per unit area of the insulating layer ($= \varepsilon_i \varepsilon_0 / d_i$), $V$ is the applied voltage, $\varepsilon_s$ is the permittivity of the semiconductor, $\varepsilon_0$ is the vacuum permittivity, $q$ is the elementary charge and $N$ is the bulk doping density of the semiconductor. This model works well outside of flat-band[16]. The Young-Lippmann equation can thus be modified to include the presence of the depletion layer in the semiconductor to give the following modified Young-Lippmann equation which describes electrowetting for a liquid-insulator-semiconductor *under depletion conditions*:

$$\cos\theta(V) = \cos\theta^0 + \frac{C_i V^2}{2\gamma\sqrt{1 \pm (2C_i^2 V / \varepsilon_s \varepsilon_0 q N)}}$$

This simple model is used to generate Supplementary Figures 4 and 5:

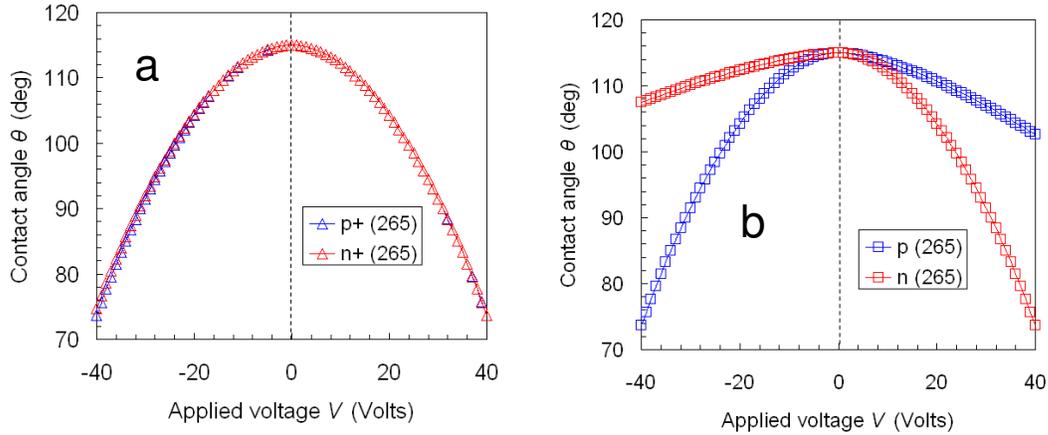

**Supplementary Figure 4**: Variation of capacitance with applied bias for the LIS system. (a) highly doped silicon wafers coated with 265 nm of Teflon®  (b) lowly doped silicon wafers coated with 265 nm of Teflon®. The following values were used for the calculation: $\gamma = 72.8$ mJ m$^{-2}$, $\theta_0 = 115°$, $\varepsilon_s = 11.9$, $\varepsilon_i = 1.92$. The n and p type doping densities were those of the commercial wafers (see **Supplementary Table I**).

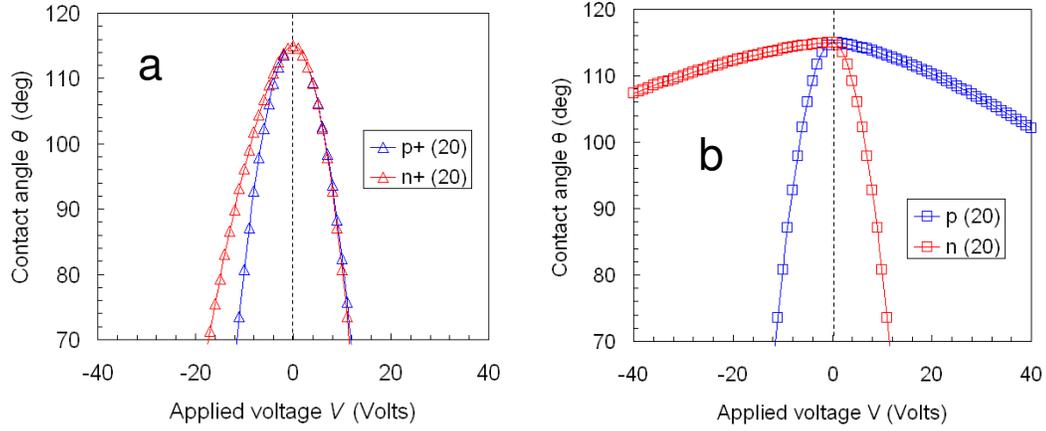

**Supplementary Figure 5**: Variation of capacitance with applied bias for the LIS system. (a) highly doped silicon wafers coated with 20 nm of Teflon® (b) lowly doped silicon wafers coated with 20 nm of Teflon®. The following values were used for the calculation: $\gamma = 72.8$ mJ m$^{-2}$, $\theta_0 = 115°$, $\varepsilon_s = 11.9$, $\varepsilon_i = 1.92$. The n and p type doping densities were those of the commercial wafers (see **Supplementary Table I**).

## 4. Model for photoelectrowetting under illumination

An expression can be formulated for the relationship between the photoelectrowetting contact angle $\theta_L$ as a function of electron-hole generation rate $g$ in the semiconductor under illumination. By substituting the expression for the photocapacitance of a MIS stack under illumination[25] into the Young-Lippmann equation one obtains:

$$\cos\theta_L(V,g) = \cos\theta^0 + \frac{V^2}{2\gamma}\left[\frac{1}{C_i} + \frac{1}{C_L}\left(\frac{\omega^2\varphi^2}{1+\omega^2\varphi^2}\right)\right]^{-1}$$

where $\varphi$ is given by the following equation[25]:

$$\varphi = \frac{N\tau}{n_i + g\tau}$$

and the limit capacitance $C_L$ of the space charge layer is given by[16]:

$$\frac{1}{C_L} = \sqrt{\frac{4kT\ln(N/n_i)}{\varepsilon_s\varepsilon_0 q^2 N}}$$

$g$ is the electron-hole generation rate (cm$^{-3}$ s$^{-1}$) in the semiconductor due to illumination with above band-gap light, $\omega$ is the angular frequency (rad s$^{-1}$), $k$ is Boltzmann's constant, $q$ is the elementary charge, $T$ is the temperature, $\varepsilon_s$ is the dielectric constant of the semiconductor (11.9 for silicon)[1], $N$ is the doping of the semiconductor, $n_i$ is the intrinsic carrier concentration of the semiconductor (1.45×10$^{10}$ cm$^{-3}$ for silicon at 300K)[16], $\tau$ is the minority carrier lifetime in the semiconductor (2.5 ms for silicon)[16]. Grosvalet and Jund[25] state that providing the measurement frequency $\omega$ is greater than $\varphi^{-1}$ (as is the case here) then the capacity is dependent solely on the variation of the space-charge limit capacitance with illumination. Under intense illumination (with above band-gap radiation)[25] $g\tau \gg n_i$ thus $\varphi \sim N/g$ then the capacitance (per surface) will be equal to $\sim C_i$ under intense illumination.

The following figures show calculations for the variation of capacitance and contact angle as a function of electro-hole pair generation rate in the silicon:

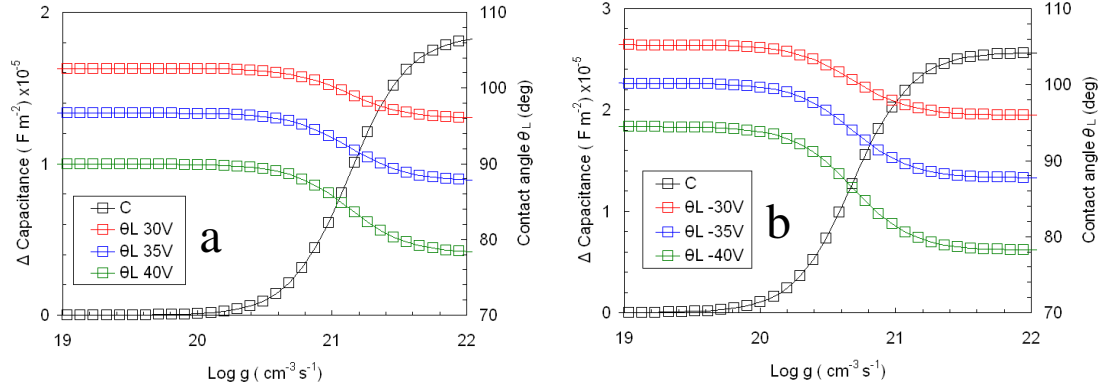

**Supplementary Figure 6**: Variation of contact angle and capacitance with electron-hole pair generation rate $g$ for the LIS system under steady-state illumination. (a) p-type silicon wafer coated with 265 nm of Teflon® (b) n-type silicon wafer coated with 265 nm of Teflon®. The following values were used for the calculation: $\gamma = 72.8$ mJ m$^{-2}$, $\theta_0 = 120°$, $t_i = 265$ nm, $\varepsilon_s = 11.9$, $\varepsilon_i = 1.92$, minority carrier concentration $n_i = 1.45 \times 10^{16}$ cm-3, minority carrier lifetime $t = 2.5 \times 10^{-3}$ s. The n and p type doping densities were those of the commercial wafers (see **Supplementary Table I**).

The change in overall capacitance when illuminating the LIS stack (at a fixed voltage $V$) is given by the following equation:

$$\Delta C = \frac{2\gamma}{V^2}\left[\cos(\theta_{light}) - \cos(\theta_{dark})\right] \qquad \textbf{(Illumination)}$$

which can be compared to the rearranged Young-Lippmann equation when applying a voltage $V$ ($C$ is fixed) for the LIC stack:

$$V^2 = \frac{2\gamma}{C}\left[\cos(\theta_V) - \cos(\theta_0)\right] \qquad \textbf{(Voltage)}$$

### 5. Measurement ramp rate and C-V evolution for a MIS.

It is well known that if the applied voltage is ramped quickly during a high frequency capacitance-voltage measurement that the structure is not in thermal equilibrium[16]. When ramping the voltage from flat-band to threshold and beyond, the inversion layer is not formed as the generation of minority carriers can not keep up with the amount needed to form the inversion layer. The depletion layer therefore keeps increasing beyond its maximum thermal equilibrium value; the result of this is a capacitance which further decreases with voltage. The

minimum ramp rate required to observe deep depletion is then obtained from the following equation:

$$\frac{dV}{dt} > \frac{qn_i x_d}{2\tau C_i}$$

using $n_i = 1.45 \times 10^{16}$ m$^{-3}$, $x_d \sim 1 \times 10^{-5}$ m, $C_i \sim 6.41 \times 10^{-5}$ F m$^{-2}$ and $\tau = 2.5 \times 10^{-3}$ s one can estimate that a voltage ramp rate > 72 mV s$^{-1}$ will result is deep depletion and the inversion capacitance will not be observed. During the electrowetting experiments (and indeed electrowetting experiments in general if a DC voltage is used[12-15]) the applied voltage is ramped very quickly (in steps) thus the electrowetting behaviour observed here will not be due to the low frequency capacitance where inversion is observed. The long life-time of the two lowly doped silicon wafers[S8] means that the deep depletion effect is observed at room temperature.

[S8]Yablonovitch, E. & Gmitter, T. Auger recombination in silicon at *low* carrier densities. *Appl. Phys. Lett.* **49**, 587-9 (1986).

**6. Breakdown in the LIS system**

The breakdown fields for Teflon® AF and Silicon are $2 \times 10^6$ V cm$^{-1}$ (Refs. 40 and S9) and $3 \times 10^5$ V cm$^{-1}$ (Ref. 16). As the capacitances can be evaluated from the C-V measurements then the fields in the Teflon® and silicon space-charge regions can be estimated as a function of applied voltage. A simple calculation of two capacitors in series (depletion mode for the n and p type silicon wafers coated with 265 nm of Teflon®) shows that at a voltage of ±40V (dark) the breakdown fields (both in the Teflon and the silicon) are not exceeded. However, in the presence of intense illumination (at ±40V) the field in the Teflon layer effectively increases but does not exceed the published breakdown field. In the case of the 20 nm thick Teflon® layer then 5V should cause breakdown in the Teflon® film; however measurements were possible to ±10V for the highly doped silicon wafers coated with 20 nm of Teflon® implying a higher breakdown field than published[40,S9] e.g. $\sim 5 \times 10^6$ V cm$^{-1}$.

[S9]Ding, S-J., Wang, P-F., Wan, X-G., Zhang, D. W., Wang, J-T. & Lee, W. W. Effects of thermal treatment on porous amorphous fluoropolymer film with a low dielectric constant. *Materials Science and Engineering* B **83**, 130–136 (2001).

# E. Additional Experiments.

**Capacitance-Voltage measurements for on lowly doped silicon (p and n) coated with 20 nm of Teflon®**

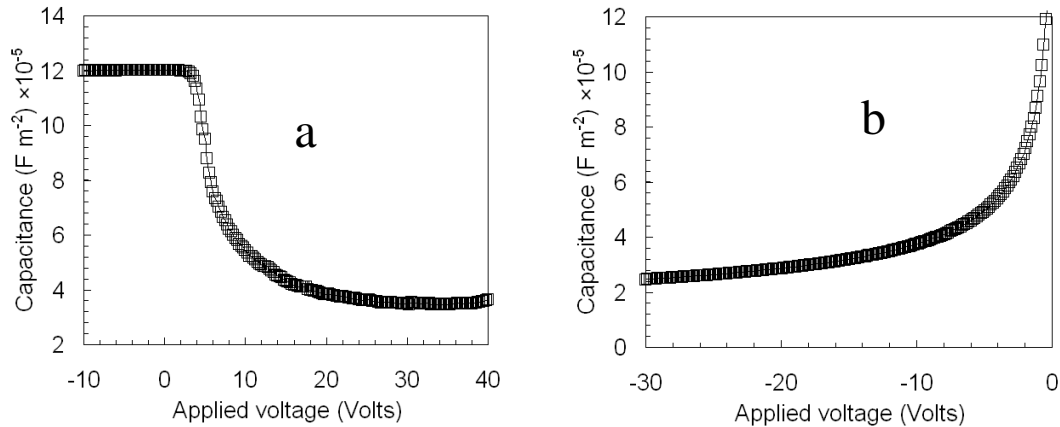

**Supplementary Figure 7**: Capacitance-Voltage measurements for (a) p-type silicon coated with 250 nm of thermally grown $SiO_2$ and 20 nm of Teflon® and (b) n-type silicon coated with 20 nm of Teflon®.

**Electrowetting on highly doped silicon wafers (p+ and n+) coated with 20 nm of Teflon®.**

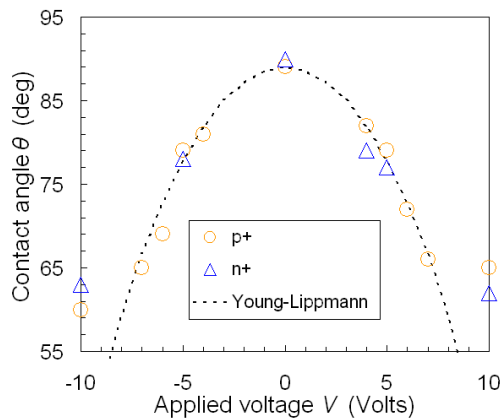

**Supplementary Figure 8**: Electrowetting on p+ and n+ silicon coated with 20 nm of Teflon® using a 0.01M NaCl solution

**Electrowetting and photoelectrowetting on bare silicon:** *Schottky electrowetting*?

Photoelectrowetting experiments were carried out by placing a conducting liquid on the surface of a bare silicon wafer, i.e. in the absence of an insulator[28,29,32]. An n-type silicon wafer was used (see Table I) which was cleaned in the same way as described in Section B above.

When applying a negative voltage to the droplet the contact angle varied little from 0 to -20V. At low voltages (0-5V) the droplet was observed to spread out upon illumination using the white light source (see above). However, the photoelectrowetting effect was not reversible as with the Teflon® coated samples. The following photographs and graph summarize the results.

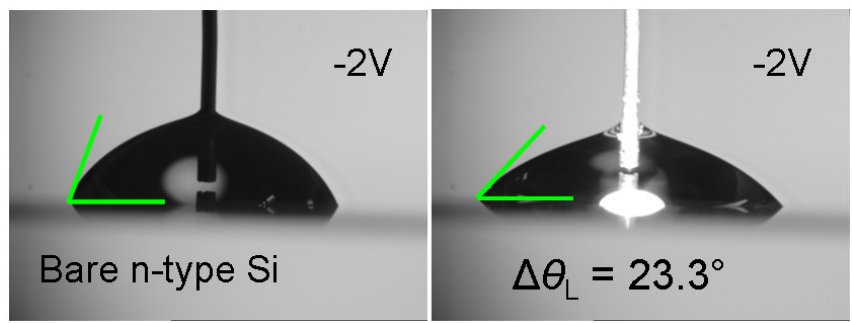

**Supplementary Figure 9**: A non-reversible photoelectrowetting effect for a conducting liquid (0.01M HCl) on a base silicon (n-type) surface at -2V.

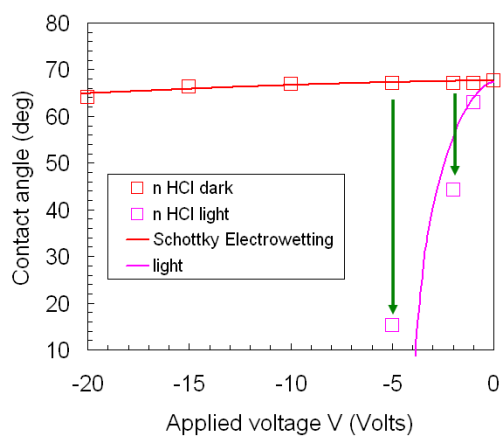

**Supplementary Figure 10**: Schottky electrowetting and photoelectrowetting for a liquid-semiconductor (LS) stack: a drop of conducting liquid (0.01M HCl solution here) is placed on the surface of a semiconductor (n-type silicon here). Evolution of contact angle in ambient room light (red squares) and under illumination (pink squares). The red line is based on the Schottky barrier model[16] to modify the Young-Lippmann equation.

It is known that a Schottky barrier[16] forms when an electrolyte is in contact with a semiconductor[28,29,32], illumination generates electron-hole pairs in this depletion zone resulting in a capacitance increase[25], the effect of this is that the contact angle reduces. The depletion zone under the electrolyte-semiconductor interface[28,29,32] can be modelled by the following equation:

$$d_s(V) = \sqrt{\frac{2\varepsilon_s \varepsilon_0 V}{qN}}$$

where $\varepsilon_s$ is the low frequency dielectric constant of the semiconductor, $q$ is the elementary charge and $N$ is the bulk doping density of the semiconductor.

Thus we have the following expression for the relationship between the contact angle and the applied voltage for a liquid-semiconductor (LS) system in reverse bias[16] (positive voltage on a p-type semiconductor or negative voltage on an n-type semiconductor):

$$\cos\theta(V) = \cos\theta_0 + \frac{\varepsilon_s \varepsilon_0}{2\gamma\sqrt{\frac{2\varepsilon_r \varepsilon_0 V}{qN}}} V^2$$

The red line in Supplementary Figure 10 is based on this equation. The pink line in Figure 10 is base on the Young-Lippmann equation with an insulator (depletion width) thickness equal to 25 nm.

## F. List of Supplementary Films

**Film 1**: n-type silicon wafer ($N = 6\times10^{14}$ cm$^{-3}$) coated with 265 nm of Teflon®. Droplet = HCl solution (0.01M).

**Film 2**: n-type silicon wafer ($N = 6\times10^{14}$ cm$^{-3}$) coated with 265 nm of Teflon®. Droplet = NaCl solution (0.001M).

**Film 3**: n-type silicon wafer ($N = 6\times10^{14}$ cm$^{-3}$) coated with 265 nm of Teflon®. Droplet = Coca-Cola Zero®.